\documentstyle[aps,prl,multicol,epsf]{revtex}
\draft

\begin{document}
\title{Quantum dots with even number of electrons: Kondo effect in a finite
magnetic field}
\author{Michael Pustilnik$^{a,b}$, Yshai Avishai$^{c}$, and Konstantin Kikoin$^{c}$,}
\address{$^{a}$Danish Institute of Fundamental Metrology,\\
Anker Engelunds Vej 1, Building 307, DK-2800 Lyngby, Denmark \\
$^{b}$ \O rsted Laboratory, Niels Bohr Institute, \\
Universitetsparken 5, DK-2100 Copenhagen, Denmark \\
$^{c}$Department of Physics, Ben Gurion University, \\
Beer Sheva 84105, Israel
\vspace{3mm}}
\date{\today}
\maketitle

\begin{abstract}
We study a small spin-degenerate quantum dot with even number of electrons,
weakly connected by point contacts to the metallic electrodes, and subject
to an external magnetic field. If the Zeeman energy $B$ is equal to the
single-particle level spacing $\Delta $ in the dot, the system exhibits
Kondo effect, despite the fact that $B$ exceeds by far the Kondo temperature 
$T_{K}$. A possible realization of this in tunneling experiments is
discussed.
\end{abstract}


\begin{multicols}{2}

Zero-bias anomaly of tunneling conductance, discovered in the early sixties,
has been explained in terms of scattering by magnetic impurities located in
the insulating layer of the tunneling junction \cite{AA}, in close analogy
with the Kondo explanation of the resistivity minimum in metals \cite{Kondo}%
. Recently, this problem gained renewed attention following theoretical
predictions that very similar effects should be detectable in tunneling of
electrons through small semiconductor quantum dots \cite{AM},\cite{MWL}. It
was indeed observed in quantum dots formed in GaAs/AlGaAs heterostructures
by the gate-depletion technique \cite{E1}- \cite{E4}.

In a quantum dot, a finite number ${\cal N}$ of electrons is confined in a
small region of space. The electrostatic potential of the dot can be tuned
with the help of capacitively coupled gate electrode. By varying the gate
voltage, one can switch between Coulomb blockade valleys, where an addition
or removal of a single electron to the dot is associated with large charging
energy $E_{c}$. In this regime, fluctuations of charge are suppressed, and $%
{\cal N}$ is a well defined integer, either even or odd. Transport, however,
is still possible by means of virtual transitions via excited states of the
dot (this mechanism is known as cotunneling). If ${\cal N}={\rm odd}$, the
dot has non-zero total spin, and the cotunneling can be viewed as a magnetic
exchange. As a result, the tunneling cross-section is expected to approach
the unitary limit at low temperature $T\ll T_{K}$ \cite{AM}. At finite
values of the source-drain voltage $eV\gg T_{K}$ the Kondo effect is
suppressed \cite{MWL}. Therefore, the width of the peak of differential
conductance at zero bias is of the order of $T_{K}$. If a magnetic field is
applied to the system, the zero-bias peak splits into two peaks at $eV=\pm B$%
, where $B=g\mu _{B}B_{\parallel }$ is the Zeeman energy \cite{MagnField}.
These peaks are observable even at $eV,B\gg T_{K}$ \cite{MWL},\cite{E2},\cite
{E3}. However, due to a nonequilibrium-induced decoherence, these peaks are
wider than $T_{K}$, even at zero temperature, and the value of the
conductance at the peaks never reaches the unitary limit \cite{MWL},\cite{WM}%
,\cite{KNG}.

For ${\cal N}={\rm even}$ this consideration is inapplicable, since in the
ground state of the {\it spin-degenerate} quantum dot all single-particle
energy levels are occupied by pairs of electrons with opposite spins, and
the total spin is zero. Therefore, the Kondo physics is not expected to
emerge in this case. Not surprisingly, investigations of dots with ${\cal N}=%
{\rm odd}$ appear more attractive. Yet, as we demonstrate below, quantum
dots with ${\cal N}={\rm even}$ may exhibit a generic Kondo effect at
certain value of the Zeeman energy $B\gg T_{K}$.

\begin{figure}[tbp]
\centerline{\epsfxsize=7.5cm \epsfbox{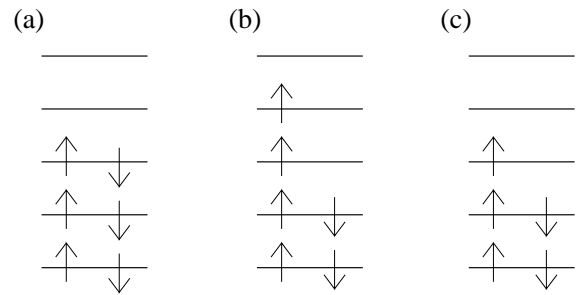}\vspace{0.6cm}}
\caption{ 
(a) The spinless ground state of the dot with ${\cal N}={\rm even}$
electrons. (b) Excited state which has $S^{z}=1$. States (a) and (b) differ
by adding a spin-down or spin-up electron accordingly to the state $\left|
\Omega \right\rangle $ of ${\cal N}-1$ electrons in the dot, shown at (c).
The states (a) and (b) are denoted as $\left| \downarrow \right\rangle $ and 
$\left| \uparrow \right\rangle $ in (\ref{states}). 
}
\label{statesFig}
\end{figure}

\begin{figure}[tbp]
\centerline{\epsfxsize=7.8cm \epsfbox{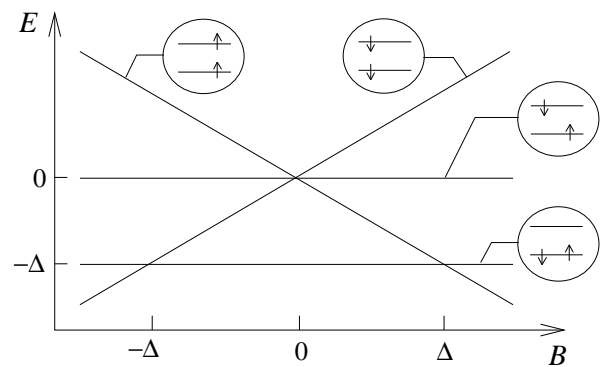}\vspace{0.7cm}}
\caption{ 
Low-energy states of a spin-degenerate quantum dot in magnetic field. 
}
\label{spectrumMF}
\end{figure}

In quantum dots, charge and spin excitations are controlled by two energy
scales, $E_{c}$ and $\Delta $ respectively, which typically differ by the
order of magnitude \cite{numbers}. This separation of the energy scales
allows to change the spin state of the dot, without changing its charge.
Indeed, if ${\cal N}={\rm even}$, the ground state of the dot has spin $S=0$%
. The lowest excited state with non-zero spin $S=1$, has energy $\Delta $
(see Fig.\ref{statesFig}). These two states, however, are affected very
differently by the magnetic field, and for $B=\Delta $, they become
degenerate (see Fig.\ref{spectrumMF}). Since they differ by flipping the
spin of a single electron in the dot, this system is a natural candidate for
realizing the Kondo effect. Moreover, this Kondo effect is unique in a sense
that the presence of a large magnetic field $B=\Delta \gg T_{K}$ is a
necessary condition for its very occurrence.

To check this idea, we consider the following model Hamiltonian for a
quantum dot, such as studied in \cite{E1}-\cite{E4}, attached to a {\it %
single} metallic electrode: 
\begin{equation}
{\cal H}=H_{0}+H_{d}+H_{T}.  \label{H}
\end{equation}
Here the Hamiltonian of the lead electrons is 
\begin{equation}
H_{0}=\sum_{k\sigma }\epsilon _{k}\psi _{k\sigma }^{\dagger }\psi _{k\sigma
},  \label{lead}
\end{equation}
where $\epsilon _{k}$ is the energy, measured from the Fermi level $\epsilon
_{F}$. An in-plane magnetic field has no influence on the two-dimensional
electron gas in the lead, provided that $B$ is small compared to the Fermi
energy (which is certainly the case for $B\sim \Delta $) \cite{MagnField}.
As for the dot Hamiltonian $H_{d}$ we consider just the two single-particle
energy levels, closest to $\epsilon _{F}$ (Note that the Anderson model
description of the ${\cal N}={\rm odd}$ case \cite{AM},\cite{MWL},\cite{WM}
is based on a similar approximation, which is valid, provided that the
conductances of the tunneling junctions are small \cite{GHL}). Hence, 
\begin{equation}
H_{d}=\sum_{p\sigma }\frac{1}{2}\left( p\Delta -\sigma B\right) d_{p\sigma
}^{\dagger }d_{p\sigma }+E_{c}\left( N-2\right) ^{2},  \label{dot}
\end{equation}
where $N=\sum_{p\sigma }d_{p\sigma }^{\dagger }d_{p\sigma }$, $p=\pm 1$
refers to single-particle energy levels in the dot, $\sigma =\pm 1$ stands
for up- and down- spin. In writing the interaction term in (\ref{dot}) we
assumed that the system is tuned to the middle of the ${\cal N}={\rm even}$
valley of the Coulomb blockade. The coupling between the dot and the
electron gas is described by the tunneling Hamiltonian 
\begin{equation}
H_{T}=\sum_{p\sigma }t_{p}\psi _{\sigma }^{\dagger }d_{p\sigma }+{\rm H.c.}%
,\;\psi _{\sigma }=\frac{1}{\sqrt{L}}\sum_{k}\psi _{k\sigma },  \label{Ht}
\end{equation}
where $L$ is a normalization constant, and we have allowed an explicit
dependence of the tunneling amplitudes on $p$. The two states of the dot
which become degenerate at $B=\Delta $, are 
\begin{equation}
\left| \uparrow \right\rangle =d_{+1\uparrow }^{\dagger }\left| \Omega
\right\rangle ,\;\left| \downarrow \right\rangle =d_{-1\downarrow }^{\dagger
}\left| \Omega \right\rangle ,  \label{states}
\end{equation}
with $\left| \Omega \right\rangle =d_{-1\uparrow }^{\dagger }\left|
0\right\rangle $, in which $\left| 0\right\rangle $ is the ground state of
the dot with ${\cal N}-2$ electrons. It is useful to define spin operators,
built on the states (\ref{states}): 
\[
S^{z}=\frac{1}{2}\left( \left| \uparrow \right\rangle \left\langle \uparrow
\right| -\left| \downarrow \right\rangle \left\langle \downarrow \right|
\right) ,\;S^{+}=S^{x}+iS^{y}=\left| \uparrow \right\rangle \left\langle
\downarrow \right| .
\]
These operators act on different {\it spin} states of the dot. Since $%
E_{c}\gg \Delta $, virtual charge excitations to states with $N\neq 2$ can
be integrated out by means of a Schrieffer-Wolf transformation. The
resulting effective Hamiltonian now reads, 
\begin{equation}
H=H_{0}+H_{p}+H_{ex}.  \label{Heff}
\end{equation}
It contains a potential scattering part 
\begin{equation}
H_{p}=U_{c}\rho +U_{s}\sigma ^{z},  \label{Hp}
\end{equation}
and an anisotropic exchange interaction 
\begin{equation}
H_{ex}=J_{c}^{z}\rho S^{z}+J_{s}^{z}{\sigma }^{z}S^{z}+\frac{1}{2}J^{\perp
}\left( \sigma ^{+}S^{-}+\sigma ^{-}S^{+}\right) .  \label{Hex}
\end{equation}
The operators appearing in (\ref{Hp}-\ref{Hex}) act on the conduction
electrons at the site of the dot. They are defined as 
\begin{equation}
\begin{array}{c}
\rho =\frac{1}{2}\left( \rho _{\uparrow }+\rho _{\downarrow }\right)
,\;\sigma ^{z}=\frac{1}{2}\left( \rho _{\uparrow }-\rho _{\downarrow
}\right) ,\;\rho _{\sigma }=\psi _{\sigma }^{\dagger }\psi _{\sigma }, \\ 
\sigma ^{+}=\psi _{\uparrow }^{\dagger }\psi _{\downarrow },\;\sigma
^{-}=\psi _{\downarrow }^{\dagger }\psi _{\uparrow }.
\end{array}
\label{def}
\end{equation}
The various coefficients in (\ref{Heff}) are $%
J_{s}^{z}=2U_{s}=2\left( t_{+1}^{2}+t_{-1}^{2}\right) /E_{c}$, $J^{\perp
}=4t_{+1}t_{-1}/E_{c}$, $J_{c}^{z}=-2U_{c}=2\left(
t_{+1}^{2}-t_{-1}^{2}\right) /E_{c}$. It should be noticed that the value of 
$U_{c}$ depends strongly on the choice of intermediate states entering the
calculation. Had we included other single-particle energy levels, than the
two closest to $\epsilon _{F}$, we would have obtained some $\sim
E_{c}/\Delta $ more contributions of the same order to $U_{c}$. In addition
to the contributions, listed in (\ref{Hp}-\ref{Hex}) there also appears a
term, proportional to $S^{z}$. This term, evidently, represents a correction
to the level spacing $\Delta $, and does not lift the degeneracy of the
states (\ref{states}), if $B$ is properly adjusted (we assume that the
tunneling alone does not change the symmetry of the ground state of the dot).

In the usual Kondo effect, the spin-charge separation ensures that potential
scattering makes no contribution. In the Hamiltonian (\ref{Heff}-\ref{Hex})
the charge degrees of freedom are coupled to the spin through the term $%
J_{c}^{z}\rho S^{z}$; in addition, there appears a potential scattering in
the spin channel, $U_{s}\sigma ^{z}$. The main effect of the potential
scattering terms $H_{p}$ is to introduce small spin-dependent corrections to
the density of states \cite{BoundState}. For our purpose it is sufficient to
replace the weakly energy-dependent functions $\nu _{\sigma }$ by their
values at $\epsilon _{F}$, 
\begin{equation}
\nu _{\sigma }=\frac{\nu _{0}}{1+\left( \pi \nu _{0}U_{\sigma }\right) ^{2}}%
,\;U_{\sigma }=\frac{1}{2}\left( U_{c}+\sigma U_{s}\right) ,  \label{DOS}
\end{equation}
where $\nu _{0}$ is the (unperturbed) density of states at the Fermi level.
The effect of the remaining exchange interaction $H_{ex}$ can be studied
with the help of standard poor man's scaling procedure, which results in a
familiar set of equations \cite{PWA} 
\begin{equation}
\frac{d{\cal J}^{z}}{d\ln D}=-\left( {\cal J}^{\perp }\right) ^{2},\;\frac{d%
{\cal J}^{\perp }}{d\ln D}=-{\cal J}^{z}{\cal J}^{\perp },  \label{scaling}
\end{equation}
for the dimensionless coupling constants 
\begin{equation}
\begin{array}{c}
{\cal J}^{z}=\left( {\cal J}_{\uparrow }^{z}+{\cal J}_{\downarrow
}^{z}\right) /2, \\ 
{\cal J}_{\sigma }^{z}=\nu _{\sigma }\left( J_{s}^{z}+\sigma
J_{c}^{z}\right) ,\;{\cal J}^{\perp }=\sqrt{\nu _{\uparrow }\nu _{\downarrow
}}J^{\perp }.
\end{array}
\label{couplings}
\end{equation}
Since initially ${\cal J}^{z}\approx \nu _{0}J_{s}^{z}>0$, the solution of
equations (\ref{scaling}) flow to the strong coupling fixed point ${\cal J}%
_{\sigma }^{z},{\cal J}^{\perp }\rightarrow \infty $.

Experimentally, the properties of this system can be probed by means of
transport spectroscopy {\cite{E1}-\cite{E4}}, when the dot is connected by
tunneling junctions to the source and drain electrodes. To describe this
situation, we add an additional index $q$ ($q=R/L$ for the right/left
electrodes) to the operators, which create/annihilate conduction electrons: $%
H_{0}=\sum_{qk\sigma }\epsilon _{k}\psi _{qk\sigma }^{\dagger }\psi
_{qk\sigma }$, $H_{T}=\sum_{qp\sigma }\left( t_{qp}\psi _{q\sigma }^{\dagger
}d_{p\sigma }+{\rm H.c.}\right) $. We are interested in the contribution to
the tunneling conductance due to the Kondo effect. The potential scattering
terms do not destroy the effect, as we have seen above. Therefore, we will
ignore these terms (thereby neglecting small corrections to the densities of
states, similar to (\ref{DOS})). It is convenient to perform a canonical
transformation \cite{AM} 
\begin{equation}
a_{\sigma }=\alpha _{\sigma }\psi _{L\sigma }+\beta _{\sigma }\psi _{R\sigma
},\;c_{\sigma }=\alpha _{\sigma }\psi _{L\sigma }-\beta _{\sigma }\psi
_{R\sigma },  \label{rotation}
\end{equation}
where 
\begin{eqnarray*}
\alpha _{\uparrow } &=&t_{L,+1}/t_{\uparrow },\;\beta _{\uparrow
}=t_{R,+1}/t_{\uparrow },\; \\
\alpha _{\downarrow } &=&t_{L,-1}/t_{\downarrow },\;\beta _{\downarrow
}=t_{R,-1}/t_{\downarrow },
\end{eqnarray*}
and 
\begin{equation}
t_{\uparrow }=\sqrt{t_{L,+1}^{2}+t_{R,+1}^{2}},\;t_{\downarrow }=\sqrt{%
t_{L,-1}^{2}+t_{R,-1}^{2}}\,.  \label{ttt}
\end{equation}
Unlike in \cite{AM}, the coefficients in (\ref{rotation}) are spin-dependent
as a result of the asymmetry of the tunneling amplitudes. It turns out that
only $a_{\sigma }$ enter the interaction terms in the effective Hamiltonian
which acquire the same form, as (\ref{Heff}) with 
\begin{equation}
J_{c}^{z}=\frac{2\left( t_{\uparrow }^{2}-t_{\downarrow }^{2}\right) }{E_{c}}%
,\;J_{s}^{z}=\frac{2\left( t_{\uparrow }^{2}+t_{\downarrow }^{2}\right) }{%
E_{c}},\;J^{\perp }=\frac{4t_{\uparrow }t_{\downarrow }}{E_{c}},
\label{parameters}
\end{equation}
and with definitions of the operators analogous to (\ref{def}) (with $\psi
_{\sigma }$ replaced by $a_{\sigma }$).

In the weak coupling regime $T\gg T_{K}$ (the characteristic energy scale of
the problem - the Kondo temperature $T_{K}$ - is discussed below) the Kondo
contribution $G_{K}$ to the differential conductance can be calculated
perturbatively from (\ref{Hex}),(\ref{rotation}). The resulting expression
is lengthy, therefore we present it only for the symmetric case $%
t_{qp}=t_{q} $, when $t_{\uparrow }=t_{\downarrow }=t$ and $%
J_{s}^{z}=J^{\perp }=J=4t^{2}/E_{c}$. In this case the result can be written
in a compact form 
\begin{equation}
G_{K}=\frac{e^{2}}{\pi \hbar }g_{0}\left( \frac{3\pi ^{2}/8}{\ln ^{2}\left(
T/T_{K}\right) }\right) ,\;g_{0}=\left( \frac{2t_{L}t_{R}}{%
t_{L}^{2}+t_{R}^{2}}\right) ^{2}  \label{WeakCoupling}
\end{equation}
In the strong coupling regime $T\ll T_{K}$ the spin-flip scattering is
suppressed, and the system allows an effective Fermi-liquid description
(see, for example, \cite{Nozieres}). The zero-bias conductance then follows
immediately from the Landauer formula, 
\begin{equation}
G_{K}=\frac{e^{2}}{2\pi \hbar }\sum_{\sigma }{\cal T}_{\sigma },\;{\cal T}%
_{\sigma }=\left( 2\alpha _{\sigma }\beta _{\sigma }\right) ^{2}.  \label{Gc}
\end{equation}
In the symmetric case (\ref{Gc}) reduces to $G_{K}=\left( e^{2}/\pi \hbar
\right) g_{0}$. By virtue of the universality of the Kondo model, the two
independent parameters, $g_{0}$ and $T_{K}$, are sufficent for the
description of $G_{K}$ in the whole temperature range $T\ll \Delta $. Notice
that, due to the asymmetry of the coefficients in (\ref{rotation}), the
transmission probabilities ${\cal T}_{\sigma }$ retain the spin dependence
even in the unitary limit, unlike in the ${\cal N}={\rm odd}
$ Kondo effect \cite{AM}. This reveals itself in the spin current in
response to the applied voltage, with the corresponding spin conductance
given by $G_{K}^{S}=\left( e^{2}/2\pi \hbar \right) \sum_{\sigma }\sigma 
{\cal T}_{\sigma }\neq 0$. However, this effect might be difficult to
measure.

The role of a finite bias $eV\gg T_{K}$, and of the magnetic field's
departures from the degeneracy points $B=\pm \Delta $ is similar to that for
the case ${\cal N}={\rm odd}$ \cite{MWL},\cite{WM}: At $B=\pm \Delta $, the
conductance exhibits peaks near zero bias, whose width saturates to $T_{K}$
in the Kondo regime $T\ll $ $T_{K}$. When the degeneracy is lifted, each of
these peaks splits to two. Therefore, the peaks at $\left( B,eV\right) $
plane are located at the points, which satisfy either an equation $\left|
B-\Delta \right| \approx eV$, or $\left| B+\Delta \right| \approx eV$. For a
fixed $eV\neq 0$, these equations have four solutions for $B$.

The possibility of experimental realizations of the proposed Kondo effect
depends crucially on the value of the Kondo temperature $T_{K}$. Since for $%
\nu _{s}=\nu _{0}$ the scaling invariant \cite{PWA} $C^{2}=\left( {\cal J}%
^{z}\right) ^{2}-\left( {\cal J}^{\perp }\right) ^{2}=\left( \frac{2\nu _{0}%
}{E_{c}}\right) ^{2}\left( t_{\uparrow }^{2}-t_{\downarrow }^{2}\right)
^{2}\geq 0$, the first of the scaling equations (\ref{scaling}) can be
written as $d{\cal J}^{z}/d\ln D=C^{2}-\left( {\cal J}^{z}\right) ^{2}$,
which after integration results 
\[
\ln \frac{E_{c}}{D}=\frac{1}{2C}\ln \left( \frac{{\cal J}^{z}-C}{{\cal J}%
^{z}+C}\right) \left( \frac{{\cal J}_{0}^{z}+C}{{\cal J}_{0}^{z}-C}\right) . 
\]
Here we have taken into account that the initial bandwidth is of the order
of $E_{c}$, and ${\cal J}_{0}^{z}$ is the bare value of ${\cal J}^{z}$. The
condition ${\cal J}^{z}\left( D=T_{k}\right) \sim 1$ gives the logarithmic
estimate of the Kondo temperature. Since $C\ll 1$, one finds 
\begin{equation}
T_{K}\sim E_{c}\exp \left[ -A/{\cal J}_{0}^{z}\right] ,  \label{Tk}
\end{equation}
where $A=\frac{1}{2\lambda }\ln \left( \frac{1+\lambda }{1-\lambda }\right) $%
, $\lambda =C/{\cal J}_{0}^{z},$ and $0\leq \lambda <1$. In the isotropic
limit $\lambda \rightarrow 0$ one has $A\rightarrow 1$ and (\ref{Tk})
reduces to the usual expression $T_{K}\sim E_{c}\exp \left( -1/\nu
_{0}J_{s}^{z}\right) $. For a given ${\cal J}_{0}^{z}$ this value is
significantly higher than that corresponding to the strongly anisotropic
limit, when $\lambda \rightarrow 1$ and factor $A$ diverges as $\ln \left(
1-\lambda \right) ^{-1}$. The parameters ${\cal J}_{0}^{z}$ and $C$, which
control the Kondo temperature $T_{K}$, can be expressed in terms of the
Kondo temperatures $T_{K}^{{\cal N}\pm 1}\sim E_{c}\exp \left( -1/\nu _{0}J_{%
{\cal N}\pm 1}\right) $ for the nearby Coulomb blockade valleys with odd
number of electrons ${\cal N}\pm 1$, since the corresponding exchange
constants are given by $J_{{\cal N}-1}=4t_{\downarrow }^{2}/E_{c}$ and $J_{%
{\cal N}+1}=4t_{\uparrow }^{2}/E_{c}$:

\[
{\cal J}_{0}^{z}\approx \frac{1}{2}\left( \frac{1}{\ln E_{c}/T_{K}^{{\cal N}%
-1}}+\frac{1}{\ln E_{c}/T_{K}^{{\cal N}+1}}\right) , 
\]

\[
C\approx \frac{1}{2}\left| \frac{1}{\ln E_{c}/T_{K}^{{\cal N}-1}}-\frac{1}{%
\ln E_{c}/T_{K}^{{\cal N}+1}}\right| .
\]
From these equations and from (\ref{Tk}) follows, that 
\begin{equation}
\min T_{K}^{{\cal N}\pm 1}\lesssim T_{K}\lesssim \max T_{K}^{{\cal N}\pm 1}.
\label{inequality}
\end{equation}
That is, $T_{K}$ is intermediate between the corresponding scales for the
neighboring Coulomb blockade valleys with odd number of electrons. It 
ensures the observability of the proposed effect in the
systems, studied in \cite{E1}-\cite{E4}.

In conclusion, we argue in this Letter that spin-degenerate quantum dots
with even number of electrons exhibit Kondo effect in a finite magnetic
field, when the Zeeman energy is equal to the single-particle level spacing
in the dot, and, therefore, is much larger than the Kondo temperature. The
effect appears due to a large difference between the characteristic energy
scales for spin and charge excitations of quantum dots, and can not be
realized with the usual magnetic impurities.

{\it Acknowledgments} - We enjoyed discussions with K. Flensberg, L.
Glazman, and K. Matveev. This work was supported by the European Commission
through the contract SMT4-CT98-9030 (MP), by Kreitman fellowship (MP), and
by Israel Basic Research Fund through its Center of Excellence program (YA).

\end{multicols}

\begin{references}
\bibitem{MagnField}  Under the conditions of the experiments \cite{E1}-\cite
{E4}, the direction of the magnetic field should be in the plane of the
two-dimensional electron gas, in order to prevent the formation of the
quantum Hall states in the lead.
\bibitem{numbers}  Typical values for the systems, studied in \cite{E1}-\cite
{E4}, are $E_{c}\sim 1\;{\rm meV}$, $\Delta \sim 0.1\;{\rm meV}$.
\end{references}
\end{document}